\documentclass[aps,prc,superscriptaddress,showpacs,showkeys]{revtex4-2}
\usepackage{graphicx}
\usepackage{bm}
\usepackage{epsfig}
\usepackage{amsfonts}
\usepackage{amssymb,amscd}
\usepackage{subfigure}
\usepackage{xcolor}
\usepackage{comment}
\usepackage{amsmath}

\begin{document}

\newcommand{\EM}[1]{\textcolor{teal}{[EM: #1]}}
\newcommand{\IR}[1]{\textcolor{red}{[IR: #1]}}
\newcommand{\AV}[1]{\textcolor{orange}{[AV: #1]}}
\newcommand{\VG}[1]{\textcolor{blue}{[VG: #1]}}

\newcommand{\Rev}[1]{\textbf{\textcolor{violet}{[Review: #1]}}}

\title{Probing  axion-like particles through the gamma-ray production from cosmic-ray  
scattering in the Milky Way dark matter halo}
\author{Victor P. Goncalves}
\email{barros@ufpel.edu.br}
\affiliation{Institute of Physics and Mathematics, Federal University of Pelotas, \\
  Postal Code 354,  96010-900, Pelotas, RS, Brazil}
\affiliation{Institute of Modern Physics, Chinese Academy of Sciences,
 Lanzhou 730000, China}

\author{Emmanuel Moulin}
\email{emmanuel.moulin@cea.fr}
\affiliation{Irfu, CEA Saclay, Université Paris-Saclay, F-91191 Gif-sur-Yvette, France}

\author{Igor Reis}
\email{igorreis@ifsc.usp.br}
\email{igor.reis@cea.fr}
\affiliation{Universidade de São Paulo, Instituto de Física de São Carlos, Av. Trabalhador São Carlense 400, São Carlos, Brazil}
\affiliation{Irfu, CEA Saclay, Université Paris-Saclay, F-91191 Gif-sur-Yvette, France}

\author{Aion Viana}
\email{aion.viana@ifsc.usp.br}
\affiliation{Universidade de São Paulo, Instituto de Física de São Carlos, Av. Trabalhador São Carlense 400, São Carlos, Brazil}

\begin{abstract}
Axion-like particles (ALP) are promising candidates to comprise all the dark matter in the universe.
We investigate the  ALP couplings to photons and electrons via astrophysical measurements through the search for very-high-energy gamma rays arising from high-energy cosmic-ray scattering off ALP populating the halo of the Milky Way.
We show  that gamma-ray signals from ALP couplings to photons and electrons 
via inverse Primakoff and Compton processes respectively,
can be probed by very-high-energy ($\gtrsim$100 GeV) gamma-ray ground-based observatories,
providing an alternative and complementary avenue to probe ALP couplings in the eV mass range.
Sensitivities of current and near-future ground-based gamma-ray observatories  improves upon one order of magnitude the current constraints from gamma-ray satellite experiments for the ALP-photon couplings in the region of masses below 10$^{-9}$ GeV. Their sensitivities reached on the  ALP-electron couplings allow probing masses below 10$^{-8}$ GeV, which are lower than the masses probed in gamma-ray satellite experiments.

\end{abstract}

\maketitle
\date{\today}

\section{Introduction} 
\label{sec:intro}
Cosmological and astrophysical observations have demonstrated that dark matter (DM) constitutes about 85\% of the matter content of the universe~\cite{Planck:2018vyg}. 
Describing its underlying microphysical properties is one of the main challenges of Particle Physics~\cite{Bertone:2010zza,Feng:2010gw}. Over the last decades,  several scenarios based on Physics beyond the Standard Model (BSM) have been proposed, predicting the existence of new particles, which are expected to weakly interact with SM particles. Such an expectation is one of the motivations for the comprehensive searches for New Physics that have been performed at the Large Hadron Collider (LHC)~\cite{Boveia:2018yeb}. Another way to probe the DM properties is the study of Cosmic Ray-DM interactions, whose center-of-mass energies can be larger than those present at the LHC and can produce observable final states, such as photons and neutrinos, that can eventually be detected in gamma-ray and neutrino observatories.

Among the compelling DM candidates are axion-like particles (ALPs), which are expected to couple to the SM fields with model-dependent couplings.  
ALPs arise in BSM theories~\cite{PaolaArias_2012,PeterSvrcek_2006,PhysRevD.81.123530} 
as a pseudo-Nambu-Goldstone boson resulting
from the breaking of a U(1) symmetry.
The non-thermal production of ALPs
could provide the correct relic density via
mechanisms including misalignment, thermal inflationary production, and from the decays of heavier particles, as discussed in, for instance, Ref.~\cite{Sikivie:2006ni}. Moreover, the existence of ALPs has implications for different astrophysical phenomena, such as the transparency of the universe to high energy photons~\cite{DeAngelis:2007dqd,Mirizzi:2007hr,Simet:2007sa} and on the evolution of stars~\cite{Ayala:2014pea,Giannotti:2015kwo,MillerBertolami:2014rka}. Another important aspect is that if DM is mostly made of ALPs, our galactic halo would be one of the main sources of these particles. Among the dedicated techniques deployed to look for ALPs are light-shining-through-wall experiments, resonant cavities, axion helioscopes and haloscopes, see, for instance, extensive reviews in Refs.~\cite{Irastorza:2018dyq,RevModPhys.93.015004}.

\begin{figure}[t]
\includegraphics[scale=0.37]{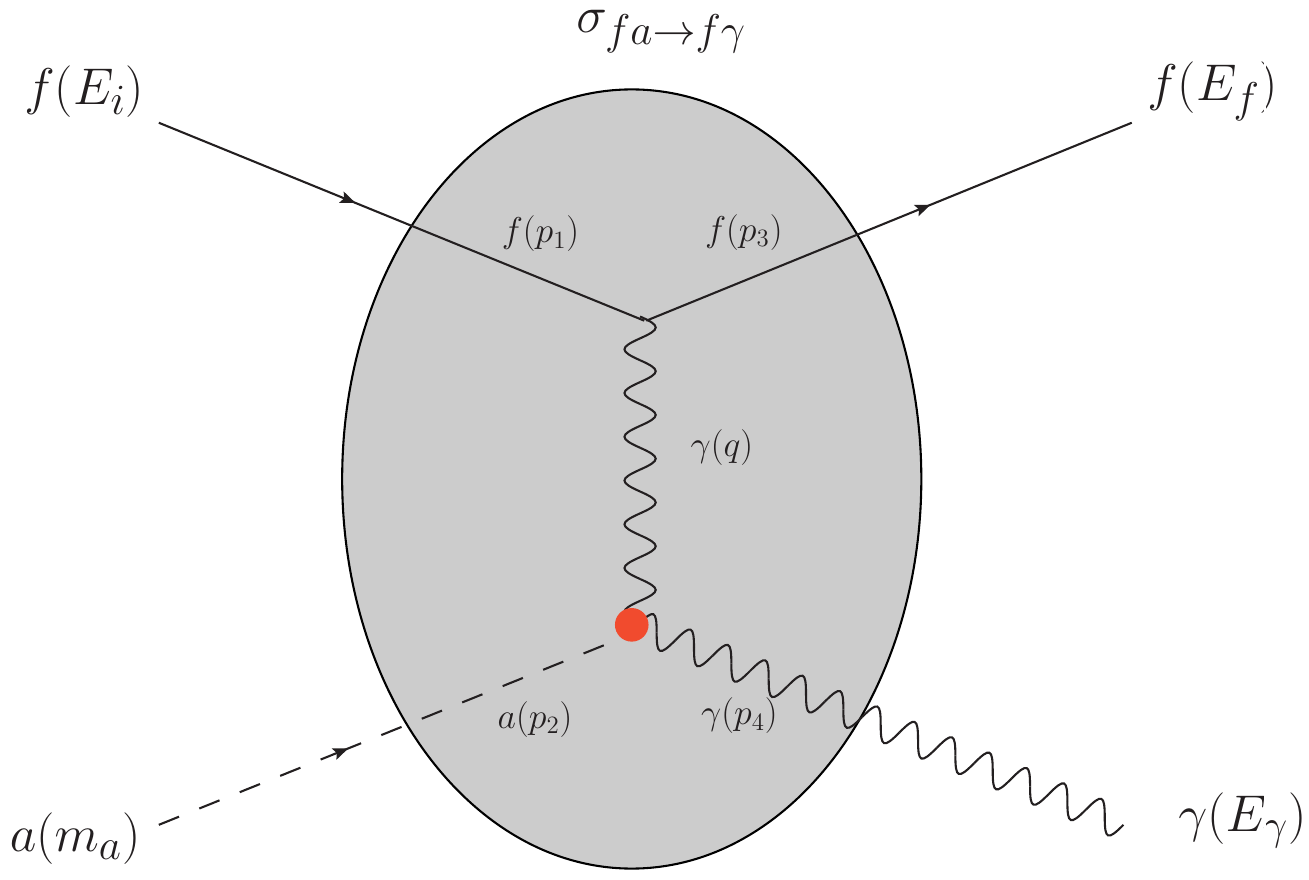}
\caption{Gamma-ray production in a CR - ALP interaction. For a photophilic ALP, the leading interaction is described by the inverse Primakoff processes, presented inside the shadowed blob. The red circle represents the ALP - photon coupling.
}
\label{Fig:diag1}
\end{figure}

In this paper, we will explore such a possibility and consider the gamma-ray production resulting from cosmic-ray CR) scattering of energetic protons and electrons on ALPs populating the Milky Way (MW) DM halo.
Our main idea, represented in Figs.~\ref{Fig:diag1} and ~\ref{Fig:diag2},  is that CRs of energy $E_{\rm i}$ colliding with an ALP $a$ of mass $m_a$,  can lose a fraction of its energy and  produce gamma rays of energy $E_{\gamma}$  in the final state. 
The associated gamma-ray flux is a function of the cross-section of the interaction between CR  and ALPs, and can be calculated given the functional form of the cosmic ray flux and the MW DM distribution. 
We show here that current and forthcoming data from the ongoing and near-future very-high-energy gamma-ray observatories such as the Cherenkov Telescope Array Observatory (CTAO) and the Southern Wide-field Gamma-ray Observatory (SWGO) can be useful to 
constrain the ALP-photon coupling in a 
region of the parameter space,
beyond the reach of current  gamma-ray satellite experiments, and provide a complementary approach to dedicated ALP experiments.

This paper is organized as follows. In the next section, we present a brief review of the formalism needed to describe the gamma-ray production in the CR-ALP interactions, considering that the primary CR can be a proton or an electron. The associated differential cross-sections are presented, as well as our assumptions for the DM density and flux computation. The main characteristics of the very-high-energy (VHE, E $\gtrsim$ 100 GeV) gamma-ray observatories considered in our analysis are presented in 
Sec.~\ref{sec:observatories} together with a discussion on the gamma-ray backgrounds, observational setup and  statistical methods used in our study. Finally, in Sec.~\ref{sec:results} we present our results for the sensitivity of the near-future ground-based gamma-ray observatories, comparing these predictions with those derived for gamma-ray satellite experiments. 

\section{ALPs and Gamma-ray production}
\label{sec:framework}
Axion-like particles are pseudo--Nambu--Goldstone bosons, which arise in models with spontaneous breaking of a global symmetry. They are expected to be characterized by a small mass compared to the scale of the spontaneous breaking and by couplings to the SM particles that are, at least, suppressed by the inverse of the same scale. Depending on the ALP mass and coupling structure, they can interact with SM gauge bosons and/or fermions. In our analysis we are particularly interested in the coupling of  the pseudoscalar ALP to  either a
photon or an electron, which is described by a Lagrangian of the form
\begin{equation}
\mathcal{L}=\frac{1}{2}\partial^\mu a \partial_\mu a -\frac{1}{2}m_a^2 a^2 -\frac{1}{4}g_{a\gamma\gamma} a F^{\mu\nu}\tilde{F}_{\mu\nu} - g_{ae} a \bar{\psi}_e \gamma_5 {\psi}_e \;,
\end{equation}
where $m_a$ is the ALP mass, $g_{a\gamma\gamma}$ is the ALP - photon coupling constant, ${F}^{\mu\nu}$ is the electromagnetic field strength tensor,  $\tilde{F}^{\mu\nu}  = \frac{1}{2}\epsilon^{\mu\nu \alpha\beta} F_{\alpha\beta}$ its dual and $g_{ae}$ is the ALP - electron coupling constant. The ALP phenomenology, as well as its experimental detection techniques, have been discussed in great details in 
reviews~\cite{Irastorza:2018dyq,Agrawal:2021dbo,RevModPhys.93.015004}.

\subsection{Proton - ALP interactions}
Initially, we will assume incident CR protons scattering off non-relativistic ALPs and focus on the gamma-ray production in the $p + a \rightarrow p + \gamma$ process, known as the inverse Primakoff process and represented in Fig.~\ref{Fig:diag1} (see, \textit{e.g.}, 
Ref.~\cite{Wu:2022psn} for a recent discussion).  Considering the ALP rest frame, one has that for a CR proton of energy $E_i = E_p$, the minimum and maximum gamma-ray energies are~\cite{Dent:2020qev}:
\begin{eqnarray}
E_{\gamma}^{\rm max/min} = \frac{m_a^2 + 2 m_{a}E_p}{2E_p + 2m_{a} \mp 2 \sqrt{E_p^2 - m_p^2}}\,\,,
\label{Eq:emin}
\end{eqnarray}
where the maximum (minimum) occurs for forward (backward) scattering. This expression implies that for very high CR energies, the maximum gamma-ray energy is close to the energy of the incoming proton. 
\begin{figure}[t]
\includegraphics[scale=0.37]{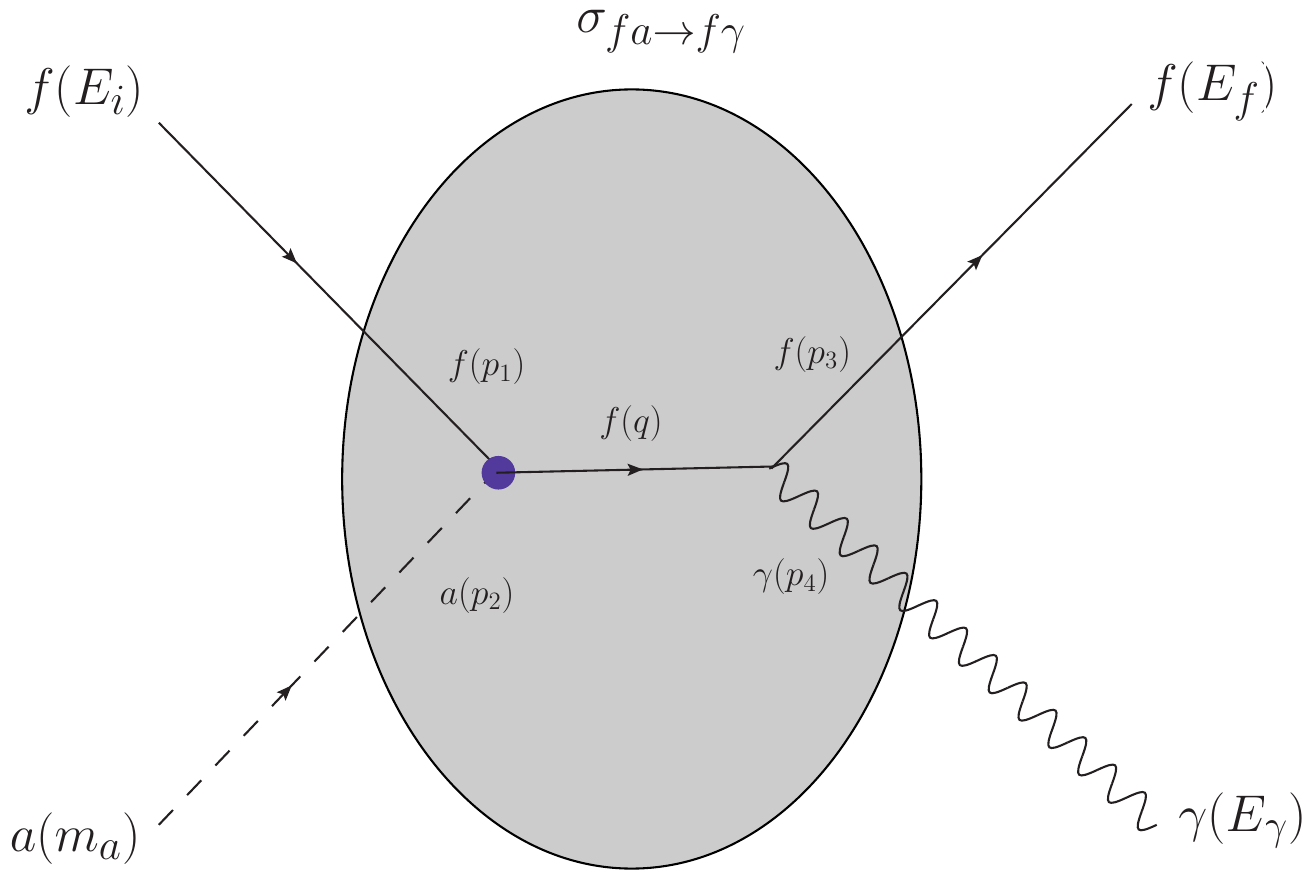}
\caption{Gamma-ray production in a CR electron - ALP interaction. For a photophobic ALP, the leading interaction is described by the inverse Compton processes, presented inside the shadowed blob. The blue circle represents the ALP - fermion coupling.}
\label{Fig:diag2}
\end{figure}

The expected gamma-ray flux from any region of the sky within a solid angle $\Delta\Omega$ is given by, \textit{e.g.}, Ref.~\cite{Dent:2020qev}:
\begin{eqnarray}
 & \,&    \frac{d\Phi_\gamma (E_{\gamma})}{dE_{\gamma}} = \frac{1}{m_{a}} \times D(\Delta\Omega)  \times  \int_{E_{p}^{\rm min}(E_{\gamma})} dE_p \frac{d\Phi_p}{dE_p}\big(E_p\big) \cdot  \frac{d\sigma_{p+a \rightarrow p + \gamma}}{dE_{\gamma}}\big(E_p,E_\gamma\big)    \, ,
    \label{eq:flux}
\end{eqnarray}
where the $D$-factor corresponds to the integral of DM density in the MW halo along  the line-of-sight  (l.o.s.) $s$ and over $\Delta\Omega$, and is given by:
\begin{eqnarray}
D(\Delta\Omega) = 
    \int_{\Delta\Omega} d\Omega\int_{l.o.s.}\rho_{\rm DM} [r(s)]ds \,.
\end{eqnarray}
The radial coordinate $r$ from the center of the halo is expressed as $r = \big(s^2 +r_{\odot}^2-2\,r_{\odot}\,s\, \cos\theta \big)^{1/2}$, where $r_\odot$ is the distance of the observer to the Galactic Center (GC), taken to be $r_\odot$ = 8.3 kpc~\cite{2020MNRAS.495.4828G}, and $\theta$ corresponding to the angle between the direction of observation and the GC.
Hence, strong signals are expected in environments with high densities of both DM and CR. The GC region is therefore an obvious target, as DM density increases towards the central region of the Milky Way, coinciding with a high VHE CR density~\cite{Fermi-LAT:2016zaq,HESS:2016pst}. In order to describe the DM density distribution in the GC region, we will assume the NFW profile parametrization given by~\cite{Navarro:1995iw},
\begin{equation}
    \rho_{\rm NFW}(r)= \frac{\rho_{s}}{(r/r_s)(1+r/r_s)^2}\,,
    \label{eq:nfw}
\end{equation}
where the profile is normalized such that $\rho(r= r_\odot) = \rho_\odot$ = 0.38 GeVcm$^{-3}$~\cite{Guo:2020oum}.
The scale radius $r_{s}$ is taken to be $r_{s}$  = 15.5 kpc~\cite{Montanari:2022buj}.
Given the dependence of the expected gamma-ray flux on the DM distribution 
through the D-factor, alternative DM profiles can be accounted for by updating the D-factor as performed in Ref.~\cite{Reis:2024wfy}.

The factor ${d\Phi_p}/{dE_p}$ in Eq. (\ref{eq:flux}) is the energy-dependent CR flux, which we assume to follow a power-law behavior 
and ${E_{p}^{\rm min}(E_{\gamma})}$ is the minimum energy of the CR proton needed to produce a gamma-ray with energy $E_{\gamma}$, which can be derived from Eq.~(\ref{Eq:emin}). 
The existence of a CR proton accelerator within the inner 10 pc of the GC has been revealed by VHE observations~\cite{Abramowski:2016mir}, exhibiting a flux with a power-law behavior given by $d\Phi_p/{dE_p}(E_{\rm p})
= \Phi_0 (E_{\rm p}/10\,\rm TeV)^{\rm -\Gamma}$.
Its slope is harder than that of local CR protons, with a normalization such that the GC proton density is approximately 6 times higher than the local CR  density. Hence, we adopt $\Phi_0 = 4 \times 10^{-8}$ cm$^{-2}$s$^{-1}$ TeV$^{-1}$sr$^{-1}$ and $\Gamma = 2.4$. Following Ref.~\cite{Abramowski:2016mir}, the baseline value of $E_{\rm p}^{\rm max}$ is taken to 1 PeV.

For an ALP of mass $m_a$, assumed to be photophilic, \textit{i.e.}, $g_{ae} = 0$, the dominant diagram is associated with the inverse Primakoff process (see, \textit{e.g.}, Refs. \cite{Wu:2022psn,Dent:2020qev}), represented by the shadowed region in Fig.~\ref{Fig:diag1}, where the red dot represents the photon-ALP coupling $g_{a\gamma \gamma}$ and $q$ is the transferred momentum of the virtual photon exchanged. For a point-like proton, one has that  ${d\sigma(p+a \rightarrow p + \gamma)}/{dE_{\gamma}}$ is given by~\cite{Dent:2020qev}:
\begin{widetext}
\begin{eqnarray}
\frac{d\sigma_{p+a \rightarrow p + \gamma}}{dE_{\gamma}} 
& = & \frac{1}{32 \pi m_a |\vec{p}|^2} \left(\frac{1}{2} \sum_{\rm spins} |{\cal{M}}|^2 \right) \,\,, \nonumber \\
& = & \frac{1}{32 \pi m_a |\vec{p}|^2} \left\{ \frac{e^2 g_{a\gamma\gamma}^2}{t^2} [m_a^2t(m_p^2 + s) - m_a^4 m_p^2 - t((s-m_p^2)^2 + st) - \frac{t(t-m_a^2)^2}{2}] \right\}\,\,,
\label{eq:alp}
\end{eqnarray}
\end{widetext}
where the factor 1/2 arises from averaging over the initial proton spin (the ALP has spin zero), $\cal{M}$ is the associated scattering amplitude to the process and $|\vec{p}|^2 = E_p^2 - m_p^2$. The Mandelstam variables $s$ and $t$ are expressed as  $s = m_p^2 + m_a^2 + 2E_p m_a$ and $t = q^2 = m_a^2 - 2E_\gamma m_a$, respectively.

\subsection{Electron - ALP interactions}
An alternative is to use CR electrons (CRe) scattering off ALPs~\cite{Sikivie:2006ni}. 
We will assume here incident local CR electrons and focus on the gamma-ray production 
via the inverse Compton process $e+a \rightarrow e+\gamma$ process (see, \textit{e.g.}, 
Refs.~\cite{Dent:2020qev,Chao:2023dwc}). In this case, we will consider that the ALP is photophilic ($g_{ae} = 0$) or photophobic ($g_{a\gamma \gamma} = 0$), with  the dominant diagrams  represented  in shadowed blobs of Figs. ~\ref{Fig:diag1} and ~\ref{Fig:diag2}, respectively. 
Due to CRe energy losses in the TeV energy region, one considers the local CRe fluxes measured by satellite experiment such as Fermi-LAT~\cite{Fermi-LAT:2017bpc}, AMS~\cite{AMS:2021nhj}, DAMPE~\cite{DAMPE:2017fbg} or CALET~\cite{CALET:2023emo}, as well as IACTs such as H.E.S.S.~\cite{HESS:2008ibn,2009A&A...508..561A} and MAGIC~\cite{Chai:2023qzm}. No electron anisotropy in their arrival direction has been detected so far~\cite{PhysRevLett.118.091103} and the CRe flux is therefore taken as isotropic. We extract the best-fit energy-differential CRe spectrum from the latest measurements from H.E.S.S.~\cite{HESS:2024etj}.
Local CRe would scatter off local DM which density is measured to be $\rho_\odot$ = 0.38 GeVcm$^{-3}$~\cite{2020MNRAS.495.4828G}. 
Given energy losses for TeV electrons in the Galactic disk~\cite{1975SSRv...17...45D}, a sphere with effective radius of $R_{\rm eff}$ = 1 kpc with constant DM density $\rho_\odot$ is assumed.
The energy-differential flux from local CRe of energy $E_e$ scattering off ALPs is given by
\begin{eqnarray}
 & \,&    \frac{d\Phi_\gamma (E_{\gamma})}{dE_{\gamma}} = \frac{\rho_\odot}{m_{a}} \times R_{\rm eff}  \times \Delta\Omega \times \int_{E_{e}^{\rm min}(E_{\gamma})} dE_e \frac{d\Phi_e^{\rm local}}{dE_e}\big(E_e\big) \cdot  \frac{d\sigma_{e+a \rightarrow e + \gamma}}{dE_{\gamma}}\big(E_e,E_\gamma\big)    \, ,
    \label{eq:CReflux}
\end{eqnarray}
where $d\Phi_e^{\rm local}/dE_e$ is the CRe flux. Following Ref.~\cite{HESS:2024etj}
, it can be well described by a smooth broken power law parametrization given by
$d\Phi_e/{dE_e}(E_{\rm e}) = \Phi_{e,0} (E_{\rm e}/1\,\rm TeV)^{\rm -\Gamma_1} [1 + (E_e/E_b)^{1/\alpha}]^{(\Gamma_1 - \Gamma_2)/\alpha}$, with flux normalization $\Phi_{e,0}$ = 1.26 $\times$ 10$^{5}$  TeV$^{-1}$ m$^{-2}$sr$^{-1}$s$^{-1}$, low-energy spectral index $\Gamma_{1} = 3.25$, high-energy power index, $\Gamma_{2} = 4.49$, break energy $E_{b}$ = 4.49 TeV, and sharpness parameter of the break $\alpha$ = 0.21.
The value of $E_{e}^{\rm min}(E_{\gamma})$ is derived using Eq.~(\ref{Eq:emin}) by replacing $E_p \rightarrow E_e$ and $m_p \rightarrow m_e$, where $m_e$ is the electron mass.

For a photophilic ALP, the differential cross-section, ${d\sigma(e+a \rightarrow e + \gamma)}/{dE_{\gamma}}$, is given by
Eq.~(\ref{eq:alp}), with the replacing of the proton mass by the electron mass. {On the other hand, for an ALP assumed to be photophobic, the dominant diagram is associated with the inverse Compton process~\cite{Dent:2020qev,Chao:2023dwc}, represented in the shadowed region of Fig.~\ref{Fig:diag2}, where $q$ is the momentum of the virtual electron and ${d\sigma(e+a \rightarrow e + \gamma)}/{dE_{\gamma}}$ is given by}~\cite{Dent:2020qev}:
\begin{widetext}
\begin{eqnarray}
\frac{d\sigma_{e+a \rightarrow e + \gamma}}{dE_{\gamma}} & = & \frac{1}{32 \pi m_a |\vec{p}|^2} \left(\frac{1}{2} \sum_{\rm spins} |{\cal{M}}|^2 \right) \,\, \nonumber \\
 & = & 
\frac{1}{32 \pi m_a |\vec{p}|^2} \Big\{e^2 g_{ae}^2\times
 \Big[-\frac{2(m_e^4 - m_e^2(2m_a^2 + s + u) + su)}{(u-m_e^2)^2}  \nonumber\\
& - & \frac{2(m_e^4 - m_e^2(2m_a^2 + s + u) + su)}{(s-m_e^2)^2} + 
    \frac{4(m_e^4 - 3m_a^2m_e^2 - m_e^2(2s + t))}{(s-m_e^2)(u-m_e^2)} + 
\frac{4(s - m_a^2)(s+t)}{(s-m_e^2)(u-m_e^2)}\Big]\Big\}\,\,,
\label{eq:alp2}
\end{eqnarray}
\end{widetext}
with $\cal{M}$ being the associated scattering amplitude, $|p|^{2} = E_e^2 - m_e^2$, $s = m_e^2 + m_a^2 + 2E_e m_a$, $t = q^2 = m_a^2 - 2E_\gamma m_a$ and $u = 2 m_e^2 + m_a^2 - s - t$. 
These equations allow us to estimate the associated gamma-ray flux for a given set of $(m_a,g_{a \gamma \gamma})$ or $(m_a,g_{a e})$ values. Conversely, we can use the experimental data for the flux to constrain these parameters.
  
\section{Observatories and data analysis}
\label{sec:observatories}
\subsection{Very-high-energy gamma-ray observatories}
In our analysis, we will focus on one existing and two near-future gamma-ray observatories. 
H.E.S.S. is an array of five Imaging Atmospheric Cherenkov Telescopes (IACTs) located in the Khomas Highlands of Namibia~\cite{HESS}. Its location provides observations of 
the GC region under ideal conditions.

The next generation of IACTs will be the Cherenkov Telescope Array Observatory~\cite{CTAO}, a two-site observatory with a telescope array in each hemisphere. The northern array will be located in La Palma, Canary Islands, and the Southern array near Paranal, Chile. It is expected  to detect gamma rays in the energy range of 20 GeV - 300 TeV and improve the flux sensitivity over current gamma-ray observatories by up to tenfold at TeV energies. CTAO is also expected to have improved angular resolution by a factor of 2 to 3 compared to current IACT observatories, and energy resolution as low as 5\% at TeV energies.
As our observations are targeted towards the GC region, we therefore make use of the expected performance for the Southern Hemisphere array of CTAO composed of 14 medium-sized telescopes and 37 small-sized telescopes,
in its \textit{Alpha} configuration. 
The instrument response functions (IRF) are taken from 
the \textit{publicly available} \texttt{prod5-v0.1} dataset~\cite{CTAprod5} 
with the \texttt{Performance-prod5-v0.1-South-20deg} scheme optimized for 50 hour observation of a point-like source at a zenith angle of 20${^\circ}$. For completeness, we also consider the \textit{Omega} configuration expected to be composed of 4 large-sized, 24 medium-sized, and 70 small-sized telescopes, with \texttt{South\_z20\_average\_50h} IRFs extracted from  
the \textit{publicly available} \texttt{prod3b-v2}
dataset~\cite{CTAOirfprod3}.

The Southern Wide-field Gamma-ray Observatory (SWGO) is a future wide-field-of-view water Cherenkov detector currently in development and planned to be built in Chile~\cite{SWGO}. It is intended to be located at high altitudes, between 10 and 30 degrees south, and to cover an energy range from a few hundred GeV up to the PeV scale.
Here, we utilize the Strawman design and the publicly-available IRFs described in Ref.~\cite{SWGOirf}.

\subsection{Backgrounds} 
The main background for the detection of gamma rays with ground-based Cherenkov Telescopes and  wide field-of-view water Cherenkov detectors comes from the misidentification of hadrons, mostly protons and helium nuclei, and electron cosmic rays as gamma rays. 
The misidentified CR protons and electrons
are referred to hereafter as the residual background. Since no VHE CR anisotropy has been detected so far, the residual background is assumed to be spatially isotropic~\cite{Fermi-LAT:2017vjf}.
This background can be determined from blank-field observations in extragalactic regions~\cite{HESS:2016mib}, or from accurate Monte-Carlo simulations of the instrument response (see, for instance, Ref.~\cite{BERNLOHR2013171}). In what follows, the residual background is obtained using the instrument response functions (IRFs) of the observatories considered here.

Additionally, the GC region harbors different VHE gamma-ray emissions originating from a variety of astrophysical sources and processes. These include the central HESS J1745-290~\cite{Aharonian:2009zk}, the supernova remnant HESS J1745-303~\cite{Aharonian:2008gw}, a diffuse emission spatially correlated with massive clouds of the Central Molecular Zone~\cite{Abramowski:2016mir}, as well as standard VHE sources in the Galactic plane~\cite{HESS:2018pbp}.
In order to avoid a complex and challenging modelling of VHE gamma-ray emissions in a self-consistent background model in the GC region, a set of masks  can be used to cover bright and extended emissions~\cite{HESS:2022ygk}. 
In the analysis presented here we exclude a band of Galactic latitudes between $\pm$0.3$^{\circ}$.

\subsection{Observation setup}
For the case of CRp-ALP scattering, in order to estimate the expected gamma-ray signals, we will  make use of the most up-to-date H.E.S.S.-like
observations of the inner GC halo~\cite{HESS:2022ygk}, obtained through the Inner Galaxy Survey program, providing 546 hours of high quality data distributed over the inner 
few degrees of the GC.  We consider a region of interest (ROI) of 3$^\circ$ in
radius around the GC, divided into 27 concentric rings of 0.1$^\circ$ width, with a homogeneous time-exposure of 500 hours. The energy-dependent acceptance and residual
background flux for H.E.S.S. observations of the inner halo of the MW are extracted from Ref.~\cite{HESS:2022ygk}.
In the case of CTAO, we choose a circular ROI with a radius of 5$^{\circ}$ around the GC and a 500-hour flat time-exposure over the entire ROI. This region is further divided into concentric annuli of 0.1$^{\circ}$ width. 
For SWGO, we consider a time exposure of 10 years over a circular ROI with a radius of 10$^{\circ}$ around the GC. 
This ROI is further divided into concentric annuli of 0.2$^{\circ}$ width.

\begin{figure*}
    \centering
    \hspace{-1cm}
    \includegraphics[scale = 0.38]{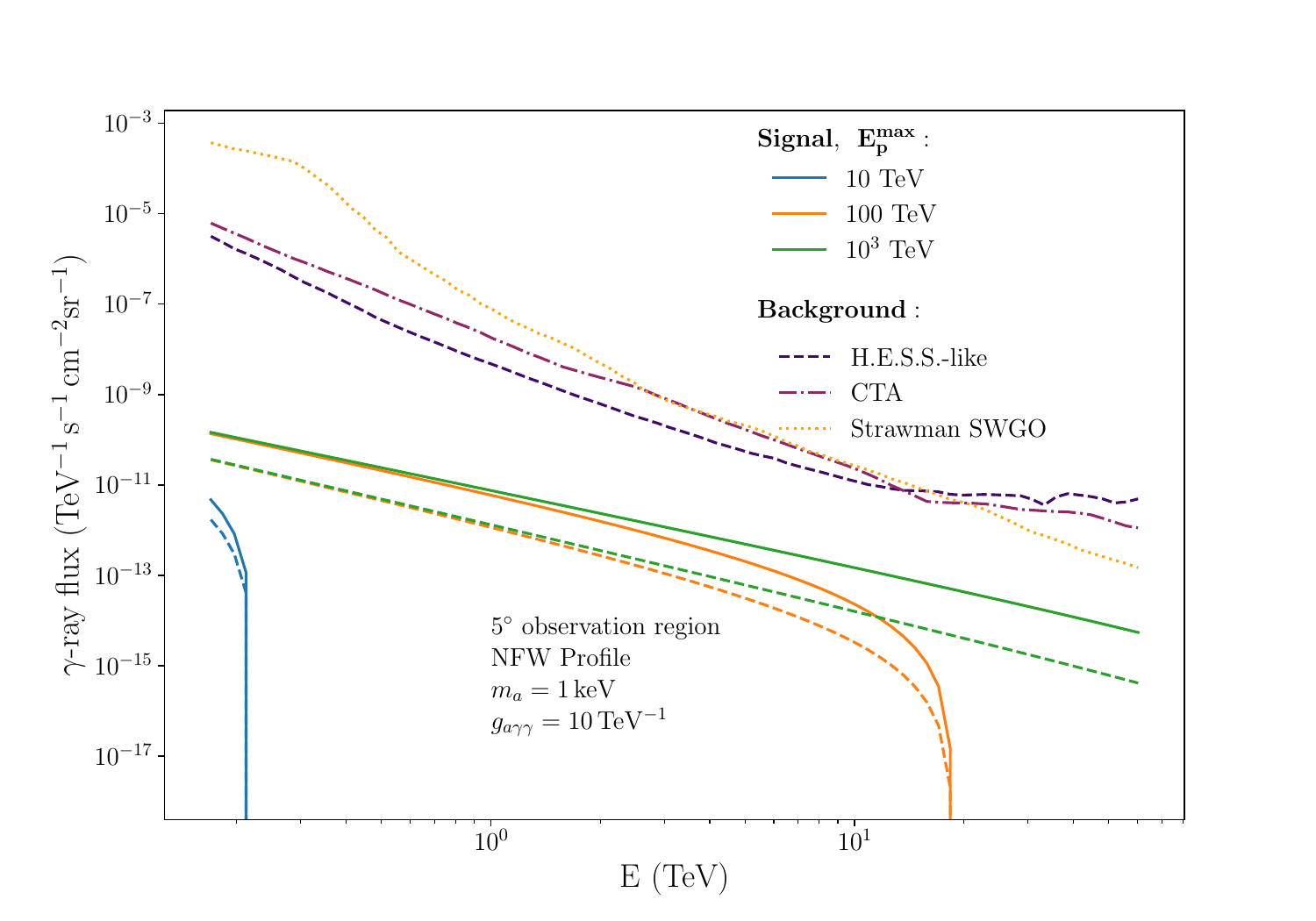}
    \hspace{-1cm}
    \includegraphics[scale = 0.38]{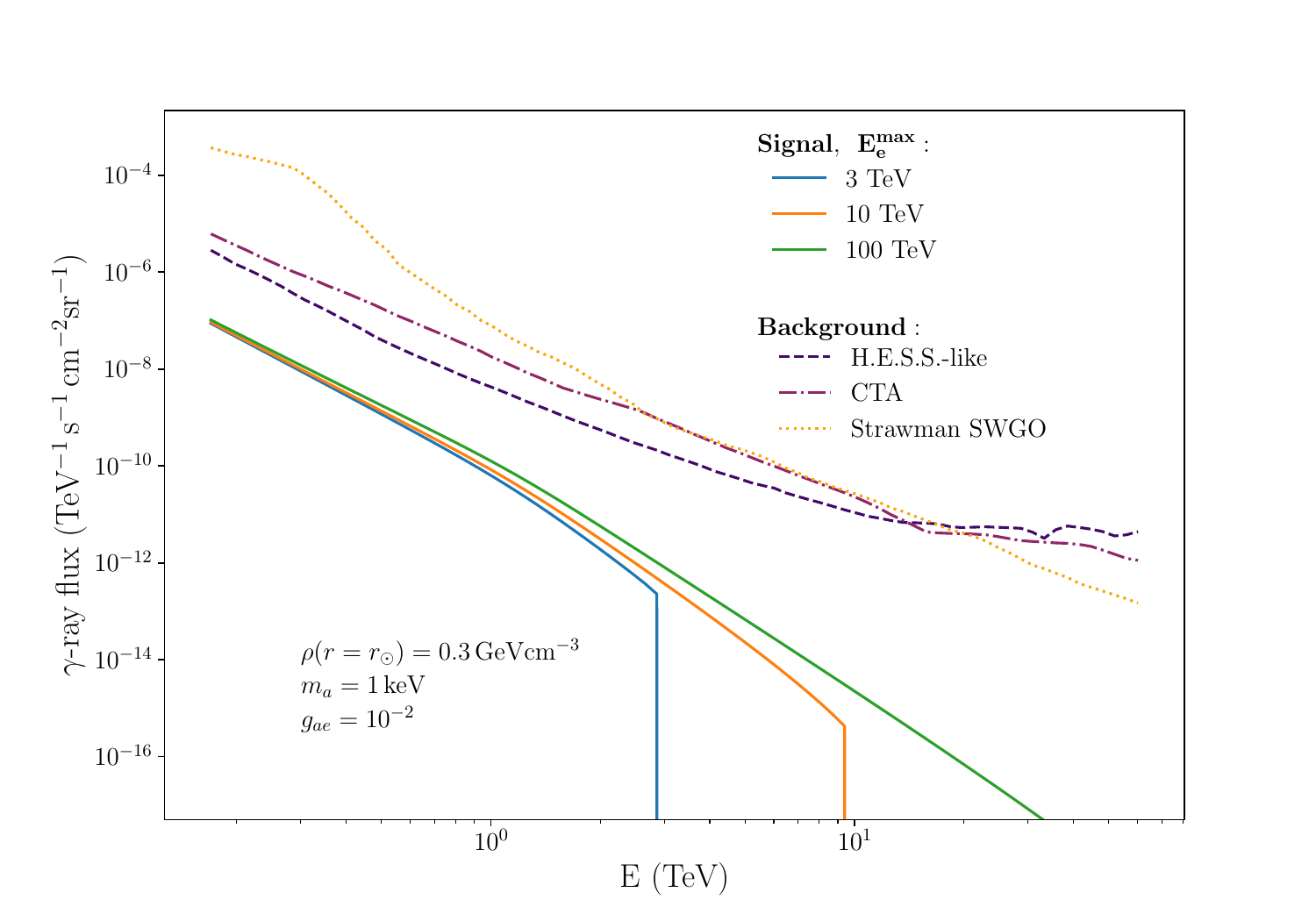} \hspace{-1.2cm}
    \caption{Energy-differential gamma-ray flux for expected CR-induced signal and background, respectively, for three observatories: H.E.S.S.-like, CTAO and SWGO for an ALP mass of 1 keV.
     {\bf Left panel:} Gamma-ray flux expected from the interaction between CR protons and ALP particles in the inner 5$^{\circ}$ of the GC region. We show expected fluxes for different maximum CR proton energies of 10 TeV (blue line), 100 TeV (orange line) and 1 PeV (green line), resespectively, as well as CRp fluxes assuming the Pevatron GC flux (solid line) and the local CR proton flux (dashed line)~\cite{Abramowski:2016mir}. For all curves we assume the NFW DM density profile,  ALP-photon $g_{a\gamma\gamma} = 10$ TeV$^{-1}$.
     {\bf Right panel:} Gamma-ray flux expected from the interaction between local CR electrons and ALPs populating the DM halo within 1 kpc of the solar neighborhood. 
    The local DM density is extracted from Ref.~\cite{2020MNRAS.495.4828G}  and the ALP-electron coupling is taken to $g_{a e} = 10^{-2}$.    
     We show the expected fluxes for maximum CR electron energies of 3 TeV (blue line), 10 TeV (orange line) and 100 TeV (green line), respectively.}
    \label{fig:fluxes}
\end{figure*}

The left panel of Fig.~\ref{fig:fluxes} shows the expected signal and background fluxes for H.E.S.S.-like observations for a DM distribution following the NFW profile, an ALP mass of 1 keV, and a coupling 
{$g_{a\gamma\gamma}$ = 10 TeV$^{-1}$}. The impact of maximum proton energy on the expected signal shows the importance of the choice of the CR accelerator together with the accessible energy range of the observatories.

For the case of CRe-ALP scattering, since the  gamma-ray signal is expected to be isotropic, we will consider extragalactic \textit{empty} fields, \textit{i.e.}, regions of the sky 
devoid of VHE gamma-ray sources and 
well outside the Galactic plane in order to avoid contamination from potential VHE Galactic diffuse emission. From recent H.E.S.S. data analysis~\cite{HESS:2024etj}, about 2800 hours of empty-field data are available with events taken in the inner 4$^\circ$ of the field of view of H.E.S.S. phase-I observations.
Therefore, we choose a region of the sky with a time-weighted solid angle of 42 h sr.
For future instruments, we will consider realistic observations to conservatively
estimate 
the total time-weighted solid angle. For CTA, a survey of empty fields 
with 6000 hours of live time is taken, which corresponds to a doubled observation time 
with respect to H.E.S.S. due to both Northern and Southern CTAO observations, with an 8$^{\circ}$ telescope field of view. 
This yields a total time-weighted solid angle of 367 h sr. As for SWGO, the live time chosen in this analysis consists of 10 years of observation, in a 1 sr region free of VHE gamma-ray sources. A total time-weighted solid angle of 21900 h sr is taken.

The right panel of Fig.~\ref{fig:fluxes} shows the expected gamma-ray signal from local CR electrons scattering off 1 keV mass ALP populating the DM halo in the Solar neighborhood assuming a coupling $g_{ae}$ = 10$^{-2}$, for H.E.S.S.-like, CTAO and SWGO observations, respectively. The impact of maximum electron energy on the expected signal shows the importance of the accessible energy range of the observatories.
As seen in Fig. \ref{fig:fluxes}, there is a difference in the flux curve shape between the proton and electron cases. For the proton-ALP interaction, the flux has a smooth cutoff as we reach the highest gamma-ray energies due to the cross-section behaviour: as the cross section approaches a given maximum proton energy, it smooths out cut-offs. In the electron-ALP interaction, however, a sharper cut-off is obtained due to the cross section increase as the electron energy approaches the chosen maximum value $\rm E_{e}^{max}$.

\subsection{Statistical method}
For a given DM mass and distribution, the expected number of signal count $N^S_{ij}$ 
in the $i$th energy and $j$th spatial bin is obtained from integrating the energy-differential gamma-ray flux, as given in Eq.~(\ref{eq:flux}),
over the solid angle $\Delta\Omega_j$ of the ROI, 
convolved with the energy-dependent gamma-ray acceptance $A_{\rm eff}^{\gamma}$ and 
energy resolution at energy $E'$, 
given by
\begin{eqnarray}
   & \, & N^S_{ij} = T_{{\rm obs}, j} \int_{E_i - \Delta E_i /2}^{E_i + \Delta E_i /2} dE  \times  \int_{-\infty}^{\infty} dE'\, \frac{d\Phi^S_{ij}}{dE'} (\Delta \Omega_j, E')\, 
    A_{\rm eff}^{\gamma}(E')\, G(E - E')\, .
\label{eq:count}
\end{eqnarray}
The energy resolution is modeled as a Gaussian $G(E - E')$, and $T_{{\rm obs}, j}$ is the time exposure in the spatial bin $j$. 
The number of expected background count $N^B_{ij}$ 
in the $i$-th energy and $j$-th spatial bins is obtained by substituting $d\Phi^S_{ij}/dE'$ by $d\Phi^B_{ij}/dE'$ in Eq.~(\ref{eq:count}). The expected background flux 
is composed of the residual background only.

We perform the statistical analysis and the computation of the expected sensitivity for each observatory following a log-likelihood ratio test statistic (TS), a well-defined and commonly used technique in gamma-ray studies (see, for instance, 
Refs.~\cite{Abdallah:2016ygi, Abdallah:2018qtu, HESS:2022ygk,Reis:2024wfy}).
The test statistics uses information on the spectral and spatial characteristics of the expected signal. We employ an ON-OFF method, where the previously defined ROIs are taken as the ON regions, and the OFF regions are taken as control regions with the same angular size as the ON regions, and are used to estimate the expected background. The likelihood used is binned in two dimensions, one energetic and one spatial, in order to compare an observation and the theoretical model, and can be written as
\begin{eqnarray}
& \, &    \mathcal{L}_{ij} (N_{ij}^S, N_{ij}^B, \bar{N}_{ij}^S, \bar{N}_{ij}^B\, |\, N_{ij}^{\rm ON}, N_{ij}^{\rm OFF})     =  \textrm{Pois}[N^S_{ij} + N^B_{ij}, N^{\rm ON}_{ij}] 
    \times  \textrm{Pois}[\bar{N}^S_{ij} + \bar{N}^B_{ij}, N^{\rm OFF}_{ij}] \, ,
\label{eq:likelihood}
\end{eqnarray}
where $\textrm{Pois}[\lambda,n] = e^{-\lambda}\lambda^n/n!$. $N^{\rm ON}_{ij}$ and $N^{\rm OFF}_{ij}$ stand for the number of observed events in the ON and OFF regions, respectively, in an energy bin $i$ and spatial bin $j$. $N^{S}_{ij}$ and $\bar{N}^{S}_{ij}$ represent the expected signal in the ON and OFF regions, respectively. Similarly, $N^{B}_{ij}$ and $\bar{N}^{B}_{ij}$ stands for the background in the ON and OFF regions. The number of background events is derived from the expected background rate and IRFs of each instrument. 
Since the misidentified CR background is obtained from Monte Carlo simulations, from here onwards we set $\bar{N}^{S}_{ij}$ = 0.

The full likelihood used in the TS can be computed by the product of the likelihood functions over all the spatial and energetic bins, such as $\mathcal{L} =  \prod_{\rm ij} \mathcal{L_{\rm ij}}$. For an assumed ALP mass $m_{a}$ and distribution, 
the amplitude of $N_{ij}^S$ is the only free parameter. Thus, this quantity is solely a function of the coupling constant $g_{a\gamma\gamma}$ or $g_{ae}$.
Therefore, for a given mass, 
the likelihood can be used to define a TS 
expressed as  
\begin{equation}
    \text{TS}(m_{a}) = - 2 \log \frac{\mathcal{L}(g, m_{a})}{\mathcal{L}(\widehat{g}, m_{a})} \, ,
    \label{eq:TS}
\end{equation}
where $\widehat{g}$ denotes the value of the coupling constant that maximizes the likelihood for a given ALP mass. This TS follows a $\chi^2$ distribution with a single degree of freedom in the limit of large statistics. 
Assuming we are working in this
limit, we can set one-sided 95\% upper limits on $g_{a\gamma\gamma}$ and $g_{ae}$ by solving for the coupling above the best fit, where TS = 2.71. Such a procedure can be used to compute upper limits given a mock dataset.

\begin{figure*}
    \centering
    \includegraphics[scale = 0.55]{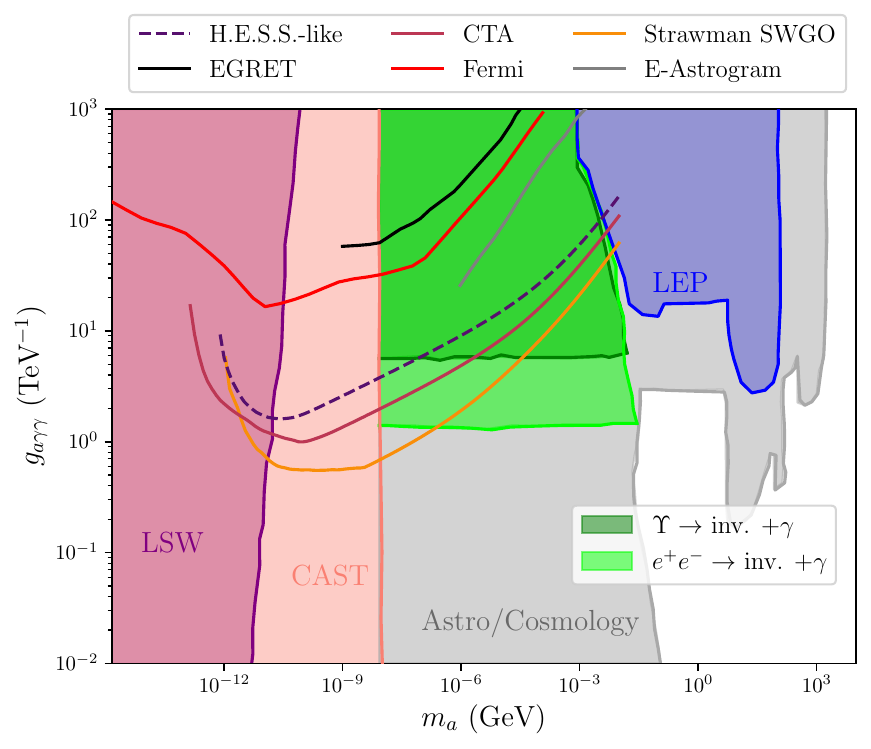}
    \includegraphics[scale = 0.55]{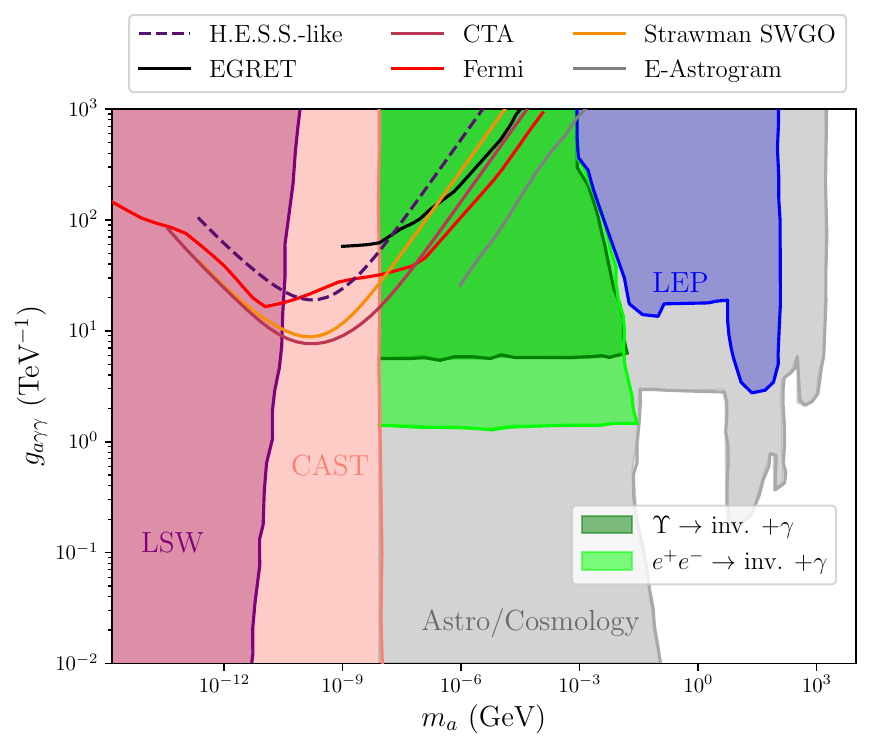}
    \caption{Sensitivity on the 
    $g_{a\gamma\gamma}$ coupling versus ALP mass $m_{\rm a}$ 
    for photophilic ALP and gamma-ray production from the 
inverse Primakoff process, shown in 
Fig.~\ref{Fig:diag1}, assuming that the incoming CR is a proton (left panel) or an electron (right panel). 
Sensitivities in the $(m_a,g_{a\gamma\gamma})$ plane are derived as 95\% C.L. mean expected upper limit for the current and near-future observatories denoted by H.E.S.S.-like, CTAO and SWGO. Limits from 
the gamma-ray based experiments Fermi, EGRET and E-Astrogram extracted from Ref.~\cite{Dent:2020qev}
are also displayed. We contrast gamma-ray results with others from 
LSW~\cite{Jaeckel:2015jla}, CAST~\cite{CAST:2017uph}, 
LEP~\cite{Jaeckel:2015jla},
rare decay and electron-positron annihilation searches \cite{Dent:2020qev}, and a combined region of different astrophysics and cosmology exclusion constraints~\cite{AxionLimits}. As for the green shaded regions, see Ref~\cite{Dolan:2017osp}. \textit{\bf Left Panel}: sensitivity curves derived considering interaction of CR protons and ALP in the Galactic center region. \textit{\bf Right Panel}: sensitivity curves derived considering the interaction of local CR electrons and ALP.}
\label{fig:results_new}
\end{figure*}
\begin{figure*}[ht!]
    \centering
\includegraphics{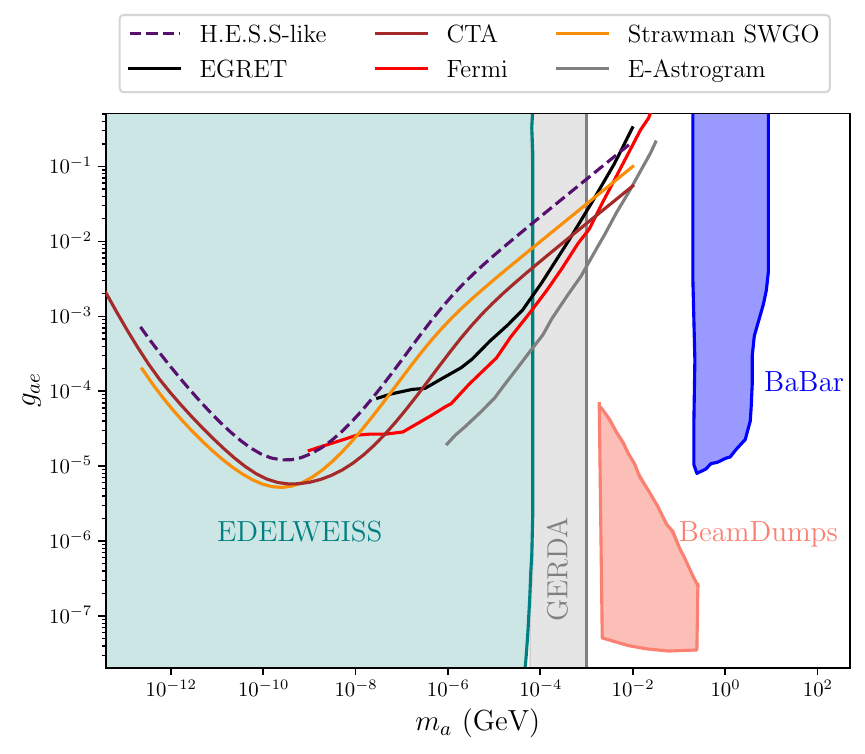}
    \caption{Sensitivity on the ALP - electron coupling $g_{ae}$ versus ALP mass $m_{\rm a}$. The sensitivity is expressed as 95\% C.L. mean expected upper limits. 
     The sensitivity constraints in the $(m_a,g_{ae})$ plane are derived considering the gamma-ray production from
    local cosmic-ray electron scattering 
     off ALPs
    for the current and near-future observatories denoted by H.E.S.S.-like, CTAO and SWGO. Exclusion regions from other experiments are also displayed.
    Current constraints from the gamma-ray  experiments Fermi, EGRET and E-Astrogram extracted from Ref.~\cite{Dent:2020qev}
    are also displayed. The gamma-ray experiment curves are shown in contrast with exclusion regions from EDELWEISS \cite{EDELWEISS:2018tde}, GERDA \cite{GERDA:2020emj}, BeamDumps \cite{Dobrich:2019dxc} and BaBar \cite{BaBar:2014zli}.}
    \label{fig:results_new_electron}
\end{figure*}

In order to derive the sensitivity of H.E.S.S.-like, CTAO and SWGO observatories, the dataset is generated using the Asimov prescription~\cite{Cowan:2011an}. 
We compute the mean expected sensitivity by treating the mean expected background as the data. The sensitivity is expressed as the mean expected upper limit, computed at 95\% C.L. . Note that this procedure  can also be used to compute confidence intervals of the expected sensitivity~\cite{Cowan:2011an}. 

\section{Results and discussions}
\label{sec:results}
Over the past decade, the search for ALPs with very small mass ($\le 1.0$ eV) has been conducted by experiments such as those considering light-shining-through-wall techniques and solar photon instruments like the CERN Axion Solar Telescope (CAST), which search for ALPs produced in the Sun. Additionally, experiments that search for ALP dark matter through electromagnetic couplings, such as ABRACADABRA  and SHAFT experiments, have been employed.
Experimental efforts are extensively reviewed in, for instance, Refs.~\cite{Irastorza:2018dyq,RevModPhys.93.015004}.
Alternatively, Fermi, EGRET and E-Astrogram gamma-ray experiments have also provided limits for the ALP-photon and ALP-electron 
couplings~\cite{AxionLimits}. 
Using the methodology described in the previous sections, we present the sensitivity reach 
of VHE gamma-ray observatories and
investigate if H.E.S.S.-like, CTAO and SWGO observatories can improve over these present bounds.

Fig.~\ref{fig:results_new} shows the sensitivity curves derived in the ($m_{a}$, $g_{a\gamma\gamma}$) parameter space
for H.E.S.S.-like, CTAO and SWGO, and compare them with exclusions from Fermi, EGRET and 
E-Astrogram, assuming the incoming CR as being a proton (left panel) or an electron (right panel). For completeness, the exclusion regions from other experiments are also 
displayed~\cite{AxionLimits}. For incident CR protons the limits obtained here are at least one order of magnitude stronger when compared to the other gamma-ray experiments. On the left panel, corresponding to the ALP-proton interaction case in the GC region, the strongest sensitivities reach 1.6 TeV$^{-1}$ for H.E.S.S.-like, 0.9 TeV${^-1}$ for CTAO and 0.5 TeV$^{-1}$ for SWGO, respectively.
Using the CR proton flux in the GC helps to significantly improve the constraints compared to the approach where the local CR electron flux is considered.
For an ALP-electron interaction in the solar neighborhood, shown in the right panel, the strongest sensitivities reach  18.7 TeV$^{-1}$ for H.E.S.S.-like, 7.6 TeV$^{-1}$ for CTAO and 8.8 TeV$^{-1}$ for SWGO. Comparing both cases, we see that $g_{a\gamma\gamma}$ is much more constrained in the ALP-proton interaction case of the inverse Primakoff process, due to the higher CR and ALP density in the GC region if compared to the local CR flux and DM density. Using a Burkert profile for the DM distribution would degrade the limits by a factor of about 4~\cite{Reis:2024wfy}.

The previous results have been derived assuming a photophilic ALP which does not couples directly with the electron. We also compute the sensitivity curves for a photophobic ALP, which allow us to derive the associated bounds for the coupling $g_{ae}$ between the electron and the ALP. Fig.~\ref{fig:results_new_electron} shows the sensitivity reach on $g_{ae}$ as a function of the ALP mass $m_{\rm a}$. Our results for the ground-based observatories are comparable to the other gamma-ray experiment limits, while probing a lower ALP mass range. The strongest constraints reach 1.2 $\times$ 10$^{-5}$ for H.E.S.S.-like, 5.8 $\times$ 10$^{-6}$ for CTAO and 5.2 $\times$ 10$^{-6}$ for SWGO. In addition, our results indicate that these observatories allow us to probe ALP-masses smaller than 1 eV.

As a summary, we have investigated here the sensitivity of high-energy astrophysical measurements to ALP-CR interactions via the production of gamma rays with current and future ground-based VHE observatories. With observations of the Galactic Center region, such observatories provide improved constraints on the $g_{a\gamma\gamma}$ coupling via CR proton-ALP interaction via the inverse 
Primakoff process with respect to satellite experiments such as EGRET, Fermi and E-Astrogram. For the CR electron case, a lower ALP mass range can be probed compared to gamma-ray satellites. Assuming a photophobic ALP, 
current and future ground-based VHE observatories are able to probe $g_{ae}$ couplings for lower ALP masses compared to gamma-ray satellite constraints via the inverse Compton process. Compared to dedicated ALP searches, astrophysical  measurements  in the VHE gamma-ray regime provide an alternative avenue to investigate ALP couplings with the strongest sensitivity achieved in the eV mass range and are complementary to searches made by experiments such as EDELWEISS, GERDA, LSW and CAST.

\begin{acknowledgments}
{The authors thank Manuel Meyer and Dieter Horns for useful discussions.}
This work was conducted in the context of the CTAO Consortium and the SWGO Collaboration. It has made use of the CTAO instrument response functions provided by the CTAO Consortium and Observatory, and the SWGO instrument response functions provided by the SWGO Consortium.
This work is supported by the "ADI 2021" project funded by the IDEX Paris ANR-11-IDEX-0003-02, and by FAPESP, process numbers 2019/14893-3, 2021/02027-0 and 2021/01089-1. A.V. is supported by CNPq grant 314955/2021-6. A.V. and I.R. acknowledge the National Laboratory for Scientific Computing (LNCC/MCTI,  Brazil) for providing HPC resources of the SDumont supercomputer (http://sdumont.lncc.br).
V.P.G. was partially supported by CNPq, FAPERGS and INCT-FNA (Process No. 464898/2014-5). 
\end{acknowledgments}

\bibliography{biblio}

\end{document}